\documentclass[aps]{revtex4}
\usepackage{amsmath}
\usepackage{amssymb}

\begin{document}

\title{Monopole-like Quantum Excitations in the Non-abelian 
Vacuum}
\author{V. Dzhunushaliev}
\email{dzhun@hotmail.kg} 
\affiliation{ Phys. Dept., Kyrgyz-Russian
Slavic University, Bishkek, Kievskaya Str. 44, 720000, Kyrgyz
Republic}

\date{\today}

\begin{abstract}
It is offered to consider monopoles in Abelian Projection 
as quantum excitations which are solutions of the quantized 
Yang-Mills equations. According to the Heisenberg quantization 
method these equations are equivalent an infinite set of equations for 
Green's functions. A procedure for cutting off these infinite series 
of differential equations after some assumptions is offered. The 
received equations are identical to equations describing a dyon. 
\end{abstract}

\maketitle

\section{Introduction}

The Abelian Projection method (for review see Ref. \cite{chernodub})
is well known in the lattice QCD simulations. 
The essence of this method is that any calculations
on the lattice with the big accuracy can be made if only field distributions
with monopoles is taken into account. The Abelian Projection allows us 
to answer on the following question : is there
monopoles in given field configuration, and says that the field 
configurations with monopoles
give an essential contribution for calculations. We would like to emphasize
that in the general case these monopoles are not dynamical 't Hooft-Polyakov
monopoles, \textit{i.e.} they are not solutions of Yang-Mills equations.
Only one property they have : a non-zero flux of magnetic field across a closed
surface. The role of Abelian Projection is to detect the presence of monopoles
in given field distribution. Every such non-dynamical monopole
(\textit{i.e.} it is not 't Hooft-Polyakov monopole) can be considered as a
quantum excitation in quantized gauge field.
\par
In this paper I would like to show that the quantization of non-abelian 
gauge field theory
leads to the appearance of a static spherically symmetric quantized field
distribution which in fact is the 't Hooft-Polyakov monopole 
and can be considered as an excitation in the non-abelian vacuum. 
In Ref. \cite{dzhsin} some arguments was presented that singular 
spherically symmetric classical solutions of 
Yang-Mills equations after quantization  give us an asymptotically well 
behaved field configuration with a flux of magnetic field. 
\par
Our main idea is that the non-abelian quantum field theory has non-perturbative
effects which can not be described on the language of Feynmann diagrams
techniques. What we can expect in this case ? Probably it 
can be (a) a static distribution of the quantized field (for example, it can
be a flux tube stretched between the quark-antiquark pair in QCD); 
(b) quantum excitations in the vacuum (for example, it can be spherically 
symmetric monopole-like quantum fluctuations of the gauge field 
in the vacuum). Such static field configurations are characterized by
$\langle A^a_\mu \neq 0\rangle$,
$\langle A^a_\mu A^b_\nu \neq 0\rangle$ and so on
(here $\langle . \rangle$ is a quantum average). These effects are
non-perturbative and the corresponding field configurations are not a
cloud of quanta.
\par
For the quantization in this case we will follow to 
Heisenberg idea \cite{heis1} \cite{heis2} . It means that we declare $A^a_\mu$
as an operator which satisfies to the corresponding 
\textbf{\textit{Yang-Mills operator equations}} \cite{dzhsin} 
\begin{equation}
  D_\nu \hat{F}^{a\mu\nu} = 0
\label{intr-1}
\end{equation}
where $\hat{F}^a_{\mu\nu} = \partial_\mu \hat{A}^a_\nu -
\partial_\nu \hat{A}^a_\mu +
f_{abc}\hat{A}^b_\mu\hat{A}^c_\nu$ is an operator of the 
field strength tensor
for the non-abelian gauge field and $f_{abc}$ is the structural constants. 
Of coarse this operator equation
has a big mathematical difficulties but following to Heisenberg
\cite{heis1} \cite{heis2}  we will apply this equation in order to obtain
an infinite set of equations connecting Green's functions of different
orders. On the language of Feynmann diagrams techniques it is the Ward
idenities connecting the Green's functions of different orders.
The operator equation \eqref{intr-1} and the above-mentioned infinite
set of equations for the Green's functions mathematically is equivalent.
The main difficulty by solving these equations for the Green's functions
is cutting off of the equations system to a finite set. It is a skill
of a physicist.
\par
Now we would like to consider some partial case : a spherically symmetric
field distribution of the quantized SU(2) non-abelian gauge field.
These words allows us to simplify Yang-Mills operator equations
\eqref{intr-1}. In order to do this we will follow by this way :
(a) on the first step we will write the spherically symmetric
ansatz for the classical SU(2) field; (b) on the
second step we will write the classical Yang-Mills equations for this
ansatz; (c) on the third step we replace the classical functions by
the quantum operators; (d) on the fourth step we will write equations
for the Green's functions and cut off the derived infinite equations
system; (e) and on the last step we will offer this field distribution 
as quantum excitation in the non-abelian vacuum. Evidently that quantum 
excitations can have an arbitrary field distribution inside itself but 
the distributions obeying the Yang-Mills equations are more 
probable. 

\section{Equations}

Now we would like to present equations describing quantum 
excitations in the vacuum of SU(2) non-abelian gauge field. 
Every such excitation can have an inner structure, \textit{i.e.} 
they can be solutions of some differential equations. In fact here 
we suppose that (according to Heisenberg) 
these equations for averaged quantum field approximately can be 
presented as classical Yang-Mills equations with quantum corrections 
coming from equations for the Green's functions of higher orders. 
The spherically symmetric ansatz is
\begin{eqnarray}
  A^a_t & = & \frac{x^a}{r^2} g(r),
\label{sec1-1a}\\
  A^a_i & = & \frac{\varepsilon_{aij}x^j}{r^2}
  \left[ 1 - f(r) \right]
\label{sec1-1b}
\end{eqnarray}
where $i=1,2,3$. Yang-Mills equations for this ansatz have the following form
\begin{eqnarray}
  r^2 f'' & = & f^3 - f g^2 - f ,
\label{sec1-3a}\\
  r^2 g'' & = & 2g f^2 .
\label{sec1-3b}
\end{eqnarray}
At first I would like to describe what we can expect from our next calculations. 
In the Ref. \cite{dzhsin2} it was shown that the general solution for the 
spherically symmetric ansatz in SU(3) non-abelian gauge theory has the 
following asymptotical behavior : $A^a_0 \propto r^\alpha$ and 
$A^a_i$ is strongly oscillating (by $r \rightarrow \infty$). In our case 
the situation is the same 
\begin{eqnarray}
  f(x) & \approx & f_0 \sin 
  \left( x^\alpha + \phi_0 \right),
\label{sec1-4a}\\
  g(x) & \approx & \alpha x^\alpha + \frac{\alpha - 1}{4} 
  \frac{\cos\left( 2 x^\alpha + 2\phi_0 \right)}{x^\alpha }
\label{sec1-4b}
\end{eqnarray}
where $x = r/r_0$ is the dimensionless radius; 
$r_0, f_0, \alpha > 1$ and $\phi_0$ are some 
constants and functions 
$f(r)$ and $g(r)$ are the classical solutions of corresponding 
Yang-Mills equations. We see that the 
frequency of the space oscillations of $f(r)$ is increased
by $r \rightarrow \infty$. But after 
the quantization the situation can be changed. The reason is that 
the quantum uncertainty $\triangle F^a_{\mu\nu}$ should obey 
the Heisenberg's uncertainty relation 
\begin{equation}
  \triangle S \approx \triangle F^2 \triangle t \triangle V 
  \approx \hbar 
\label{sec1-7}
\end{equation}
where $\triangle S$ and $\triangle F$ is uncertainties of the action $S$ 
and the non-abelian field $F^a_{\mu\nu}$ respectively; 
$\triangle V = \triangle z \triangle r \triangle l$ 
($\triangle l = r \triangle \varphi$). Let us define $\triangle r$ as the 
distance between two adjacent maxima of the function $f(r)$ 
\begin{eqnarray}
  \left( x + \triangle x \right)^2 & \approx & 
  x^2 + 2 x \triangle x \approx x^2 + 2\pi , 
  \qquad x = \frac{r}{r_0} \gg 1
\label{sec1-8a}\\
  r & \approx & \frac{2 \pi r_0^2}{\triangle r}.
\label{sec1-8b}
\end{eqnarray}
It means that inside of the small 
volume $\triangle V$ (which is 
on the distance $r \approx 2\pi r_0^2/ \triangle r$ from the origin) 
the field fluctuations can be very big 
\begin{equation}
  \triangle F^2 \triangle t \approx \frac{\hbar}
  {\triangle z \triangle r \triangle l}
\label{sec1-9}
\end{equation}
and for $\triangle r \approx 2\pi r_0^2/r$ 
\begin{equation}
  \triangle F^2 \triangle t \approx \frac{\hbar}
  {r_0^2 \triangle z \triangle l} r ,
\label{sec1-10}
\end{equation}
\textit{i.e.} there is $\tilde{r}_0$ where $\triangle F \approx F$. 
The consequence of this is that the quantization of strongly 
oscillating fields should smooth these fluctuations. But it is not all : 
the functions $f(r)$ and $g(r)$ are connected with the Yang-Mills 
equations. In Ref. \cite{dzhsin} it is shown that in the consequence of 
the field equations the bad asymptotical behavior of these functions 
$f(r)$ and $g(r)$ will 
be changed. In this paper we will try to describe this process more 
exactly. 
\par 
Following to Heisenberg we replace classical functions $f(r)$
and $g(r)$ by quantum operators $\hat{f}(r)$ and $\hat{g}(r)$
\begin{eqnarray}
  r^2 \hat f'' & = & \widehat{f^3} - \widehat{f g^2} - \hat f ,
\label{sec1-5a}\\
  r^2 \hat g'' & = & 2\widehat{g f^2} .
\label{sec1-5b}
\end{eqnarray}
Let $| Q \rangle$ is a quantum state which describes (in some
approximation) the spherically symmetric distribution of quantized
field. It means that the following quantum average exists 
$\langle f(r) \rangle = \langle Q |\hat f(r)| Q \rangle$,
$\langle f^3(r) \rangle = \langle Q |\widehat{f^3(r)}| Q \rangle$
and so on. For these Green's functions we have the following two
equations which are the first in the infinite series of equations
\begin{eqnarray}
  r^2 \left\langle f'' \right\rangle & = &
  \left\langle {f^3} \right\rangle - \left\langle f g^2 \right\rangle -
  \left\langle f \right\rangle,
\label{sec1-7a}\\
  r^2 \left\langle g'' \right\rangle & = &
  2 \left\langle g f^2 \right\rangle .
\label{sec1-7b}
\end{eqnarray}
The above-mentioned difficulty is that we must have the corresponding
equations for $\langle f^3 \rangle$, $\langle fg^2 \rangle$,
$\langle gf^2 \rangle$ and so on up to infinity. Now we would like
to cut off this infinite process.

\section{Cutting off}

Let us to introduce a new function $\varphi (r)$ which will describe
the influence of the Green's functions of higher orders in 
Eq. \eqref{sec1-7a} by such a way
\begin{equation}
  \left\langle f(r_1) g(r_2) g(r) \right\rangle =
  \left\langle f(r_1) \right\rangle
  \left\langle g(r_2) \right\rangle
  \left\langle g(r) \right\rangle -
  \left\langle f(r_1) \right\rangle \varphi(r_2) \varphi(r) .
\label{sec1-9a}
\end{equation}
In order to obtain an equation for $\varphi(r)$ we act with the operator
$r^2 \frac{d^2}{dr^2}$ on the left-hand and right-hand sides of this
definition. In the consequence we have
\begin{equation}
  \langle f(r_1) \rangle \varphi(r_2) r^2
  \varphi ''(r) =
  2\biggl(
  \left\langle f(r_1) \right\rangle
  \left\langle g(r_2) \right\rangle
  \left\langle g(r) f^2(r) \right\rangle -
  \left\langle f(r_1) g(r_2) g(r) f^2(r) \right\rangle
  \biggl).
\label{sec1-10a}
\end{equation}
But we are interesting only for the case $r_{1,2} \rightarrow r$
\begin{equation}
  \left\langle f \right\rangle \varphi r^2 \varphi '' =
  2\biggl(
  \left\langle f \right\rangle
  \left\langle g \right\rangle
  \left\langle f^2 g \right\rangle -
  \left\langle f^3 g^2 \right\rangle
  \biggl)
\label{sec1-11}
\end{equation}
here we suppose that $\hat{f}(r) \hat{g}(r) = \hat{g}(r) \hat{f}(r)$. 
It is interesting to emphasize here the absence of the $\delta-$ function. 
Heisenberg wrote in this occasion \cite{heis2} : ``The states of Hilbert 
space II 
change the commutator in such a manner that the $\delta-$ and $\delta '-$ 
functions on the light-cone disappear, and that actually the wave function 
$\psi$ and $\psi^+$ anti-commute everywhere on the subspace $t=const$. '' 
It is necessary to note that such calculations are like to Heisenberg's 
quantization of a non-linear spinor field \cite{heis1} \cite{heis2} 
and on the classical 
level to the average of the product of turbulence velocities 
$\langle v_i v_j \rangle$, $\langle v_i v_j v_k \rangle$ and so on 
\cite{landau}. 
Our main assumption is that
\begin{eqnarray}
  \left\langle f^3 g^2 \right\rangle & \approx &
  \left\langle f^2 \right\rangle
  \left\langle fg^2 \right\rangle ,
\label{sec1-12a}\\
  \left\langle f^2g \right\rangle & \approx &
  \left\langle f^2 \right\rangle
  \left\langle g \right\rangle .
\label{sec1-12b}
\end{eqnarray}
In this case we have
\begin{equation}
  \left\langle f \right\rangle \varphi r^2 \varphi'' =
  2 \left\langle f^2 \right\rangle
  \left(
  \left\langle f \right\rangle \left\langle g^2 \right\rangle
  - \left\langle f g^2 \right\rangle
  \right)
\label{sec1-13}
\end{equation}
From Eq. \eqref{sec1-9a} we see that
\begin{equation}
  \left\langle f(r) \right\rangle \varphi ^2(r) =
  \left\langle f(r) \right\rangle
  \left\langle g(r) \right\rangle ^2 -
  \left\langle f(r) g^2(r) \right\rangle .
\label{sec1-14}
\end{equation}
It leads to the following equation
\begin{equation}
  r^2 \varphi '' = 2 \varphi \left\langle f^2 \right\rangle
\label{sec1-15}
\end{equation}
and we have
\begin{eqnarray}
  r^2 \left\langle f \right\rangle '' & = &
  \left\langle f^3 \right\rangle -
  \left\langle f \right\rangle \left\langle g \right\rangle ^2 +
  \left\langle f \right\rangle \varphi ^2 -
  \left\langle f \right\rangle ,
\label{sec1-16a}\\
  r^2 \left\langle g \right\rangle '' & = &
  2 \left\langle g \right\rangle \left\langle f^2 \right\rangle ,
\label{sec1-16b}\\
  r^2 \varphi '' & = &
  2 \varphi \left\langle f^2 \right\rangle
\label{sec1-16c}
\end{eqnarray}
with the following approximations
\begin{eqnarray}
  \varphi ^2 & = & \left\langle g \right\rangle ^2 -
  \frac{\left\langle f g^2 \right\rangle}
  {\left\langle f \right\rangle} ,
\label{sec1-17a}\\
  \left\langle f^3 g^2 \right\rangle & \approx &
  \left\langle f^2 \right\rangle
  \left\langle f g^2 \right\rangle ,
\label{sec1-18b}\\
  \left\langle f^2 g \right\rangle & \approx &
  \left\langle f^2 \right\rangle
  \left\langle g \right\rangle .
\end{eqnarray}
If we suppose that
\begin{equation}
  \left\langle f^3 \right\rangle \approx
  \left\langle f \right\rangle \left\langle f^2 \right\rangle
\label{sec1-19}
\end{equation}
then the first equation will be 
\begin{equation}
  r^2 \left\langle f \right\rangle '' =
  \left\langle f \right\rangle
  \left(
  \left\langle f^2 \right\rangle - \left\langle g \right\rangle ^2 +
  \varphi ^2 - 1
  \right)
\label{sec1-20}
\end{equation}
and finally if $\langle f^2 \rangle \approx \langle f \rangle^2$ 
(it means that $\langle \Delta f^2\rangle \ll \langle f \rangle^2$ 
and fluctuations of the field $f$ is very small) 
we have the following set of equations
\begin{eqnarray}
  r^2 \left\langle f \right\rangle '' & = &
  \left\langle f \right\rangle ^3 -
  \left\langle f \right\rangle \left\langle g \right\rangle ^2 +
  \left\langle f \right\rangle \varphi ^2 -
  \left\langle f \right\rangle
\label{sec1-21a}\\
  r^2 \left\langle g \right\rangle '' & = &
  2 \left\langle g \right\rangle \left\langle f \right\rangle ^2,
\label{sec1-21b}\\
  r^2 \varphi '' & = &
  2 \varphi \left\langle f \right\rangle ^2
\label{sec1-21c}
\end{eqnarray}
It is very interesting to compare these equations with the equations
describing a dyon \cite{julia} 
\begin{eqnarray}
  r^2 f'' & = & f^3 -fg^2 + f \varphi ^2 - f ,
\label{sec1-22a}\\
  r^2 g'' & = & 2g f^2 ,
\label{sec1-22b}\\
  r^2 \varphi '' & = & 2 \varphi f^2 +
  \lambda \varphi \left( \varphi ^2 - r^2 \varphi^2_0 \right)
\label{sec1-22c}
\end{eqnarray}
where $f$ and $g$ are the same as in Eq's 
\eqref{sec1-1a} \eqref{sec1-1b} and $\varphi(r)$ describes 
the spherically symmetric Higgs scalar field $\phi^a$ 
\begin{equation}
  \phi^a (r)= \frac{x^a}{r^2} \varphi (r) .
\label{sec1-23}
\end{equation}
Immediately we see that our equations coincide with dyon equations
in the limit $\lambda \rightarrow 0$. Consequently the solution is
\begin{eqnarray}
  \left\langle f(x) \right\rangle & = &
  \frac{x}{\sinh x} ,
\label{sec1-24a}\\
  \left\langle g(x) \right\rangle & = & 
  \sinh \gamma 
  \left(
  \frac{x}{\tanh x} - 1
  \right) ,
\label{sec1-24b}\\
  \varphi (x) & = & \cosh \gamma 
  \left(
  \frac{x}{\tanh x} - 1
  \right) 
\label{sec1-24c}  
\end{eqnarray}
where $x = r/r_0$ and $\gamma$, $r_0$ are some constants. Thus we can 
say that Eq's \eqref{sec1-24a}-\eqref{sec1-24c} describe the quantized 
spherically symmetric configurations of the SU(2) gauge field with 
taking into account the higher order of Green's functions. Another words  
the function $\varphi$ is the result of the non-linearity of non-abelian 
field. It allows us to do the cautious assumption that the Higgs scalar 
field in SU(2) and SU(3) gauge theories is the consequence of the 
quantization of the non-linear fields. 
\par 
Our interpretation of the solution \eqref{sec1-24a}-\eqref{sec1-24c} 
is the following : \textbf{\textit{it describes the spherically 
symmetric excitation in the non-abelian vacuum.}} 

\section{Higgs potential}

Let us consider more carefully the right-hand side of Eq. 
\eqref{sec1-11}. We can introduce a new function $\chi(r)$ by the 
following way 
\begin{equation}
  \left\langle f \right\rangle
  \left\langle g \right\rangle
  \left\langle f^2 g \right\rangle -
  \left\langle f^3 g^2 \right\rangle = 
  \left\langle f^2 \right\rangle 
  \biggl (
  \left\langle f \right\rangle 
  \left\langle g^2 \right\rangle -
  \left\langle f g^2 \right\rangle
  \biggl )
  \left(
  1 + \frac{1}{2} \chi
  \right) = 
  \left\langle f^2 \right\rangle \left\langle f \right\rangle 
  \varphi^2 
  \left(
  1 + \frac{1}{2} \chi
  \right)
\label{sec4-10}
\end{equation}
The corresponding equation for $\varphi$ looks as 
\begin{equation}
  r^2 \varphi '' = 
  2 \varphi \left\langle f \right\rangle ^2 + 
  \varphi \left\langle f^2 \right\rangle \chi
\label{sec4-20}
\end{equation}
Of coarse, for the definition of $\chi$ we should have another 
equation from the infinite set of equations for the Green's functions. 
Such chain of differential equations have need for break. Evidently 
the next equation for the function $\chi$ will be differential equation. 
In order to avoid this differential equation we should introduce some 
algebraic relation between new function $\chi$ and old functions 
$f, g, \varphi$. For this algebraic relation 
we suppose that this addition should not destroy the 
monopole-like behavior of the initial solution (without $\chi(r)$). 
It means that $\langle f^2 \rangle \chi$ should be the Higgs potential 
\begin{equation}
  \left\langle f^2 \right\rangle \chi \approx 
  \lambda \left( \varphi ^2 - \varphi ^2_\infty \right) 
\label{sec4-30}
\end{equation}
where $\lambda$ is some constant and 
$\varphi (r) \rightarrow \varphi_\infty(r)$ by $r \rightarrow \infty$. 
More concretely we see from Eq. \eqref{sec1-24c} 
\begin{equation}
  \varphi _\infty (r) = \frac{\sinh \gamma}{r_0} r = 
  \varphi_0 r
\label{sec4-30a}
\end{equation}
where $\varphi_0 = \sinh \gamma / r_0$. Thus we have 
\begin{eqnarray}
  r^2 \left\langle f \right\rangle '' & = &
  \left\langle f \right\rangle ^3 -
  \left\langle f \right\rangle \left\langle g \right\rangle ^2 +
  \left\langle f \right\rangle \varphi ^2 -
  \left\langle f \right\rangle
\label{sec4-40a}\\
  r^2 \left\langle g \right\rangle '' & = &
  2 \left\langle g \right\rangle \left\langle f \right\rangle ^2,
\label{sec4-40b}\\
  r^2 \varphi '' & = &
  2 \varphi \left\langle f \right\rangle ^2 + 
  \lambda \varphi 
  \left(
  \varphi ^2 - \varphi^2_\infty
  \right).
\label{sec4-40c}
\end{eqnarray}
That completely coincides with the dyon equations. From equation 
\eqref{sec4-10} we see that the function $\chi$ is the next quantum 
correction for our equations, \textit{i.e.} it is the correction 
$\chi$ to the correction $\varphi$. It is necessary to note that 
Eq. \eqref{sec4-40c} is similar to the Ginzburg-Landau equation 
in the superconductivity theory. In both theories they are pure quantum 
equations connected with the non-linearity of initial classical 
equations and are the consequence of the Heisenberg quantization 
method applied for the non-abelian gauge theory and for the quantum 
solid theory \cite{dzh3}. 

\section{Discussion and Conclusions}

The above-mentioned discussion about function $\varphi (r) $ and 
an analog of Higgs potential 
$\langle f^2 \rangle \chi \approx \lambda ( \varphi^2 - \varphi^2_\infty )$ 
allows us to present such interpretation of the Higgs field : 
(a) it describes quantum corrections to the classical equations; 
(b) unstable vacuum $\varphi = 0$ describes classical singular solutions; 
(c) stable vacuum $\pm\varphi_\infty$ describes stable state 
of Yang-Mills field in the quantum region. It means that 
\textbf{\textit{the Higgs field 
describes the transition from unstable classical state to the 
quantum steady-state}}. It is possible that our interpretation of 
the function $\varphi$ 
as an analog of the Higgs scalar field is correct in the general case. 
\par 
Finally we can conclude 
\begin{enumerate}
  \item 
  Among quantum excitations (which are indicated by Abelian Projection) 
  in the non-abelian vacuum there are the dynamical monopoles. These excitations 
  exist some time which can be defined from the uncertainty principle 
  $\Delta t \approx \hbar /Mc^2 \approx 10^{-27} s$ where $M \approx 10^9 eV$ 
  is the monopole mass. 
  \item 
  They are the spherically symmetric configurations of the quantized 
  field which can be obtained from quantum Yang-Mills equations. 
  \item 
  Quantum corrections of the classical Yang-Mills equations in the first 
  approximation is connected with the Higgs scalar field. It is interesting 
  to note that if it is true then the Higgs mechanism for electro-weak 
  interactions and confinement in color chromodynamics have one origin : 
  they are quantum corrections in the consequence of the non-linearity 
  of the initial classical Yang-Mills equations.
\end{enumerate}
In Ref. \cite{dzhsin2} is shown that SU(3) gauge field theory 
has very similar classical equations for the spherically 
symmetric field configuration. It allows us to suppose 
that in the SU(3) chromodynamics quantum excitations also 
are the consequence of quantized SU(3) Yang-Mills theory. 

\section{Acknowledgment}
This work is supported by ISTC grant KR-677 and I am very grateful  
for the Alexander von Humboldt Foundation for the support of this work.

\end{document}